\documentclass{iau_FM}
\usepackage{graphicx}

\title[FM~14.~~Gravitational Wave Symphony of Structure Formation] 
{Growth of supermassive black holes, galaxy mergers and supermassive binary black holes}

\author[S. Komossa et al.]   
{Komossa, S.$^1$, 
Baker, J.G.$^2$,
 \and Liu, F.K.$^3$}

\affiliation{$^1$Max-Planck-Institut f\"ur Radioastronomie, Auf dem H\"ugel 69,
   53121 Bonn, Germany \\ email: {\tt skomossa@mpifr.de}  \\[\affilskip]
$^2$NASA/GSFC, 
    Mail Code: 663, 
    Greenbelt, MD 20771   \\[\affilskip]
$^3$Department of Astronomy, Peking University, Beijing 100871, China }  

\pubyear{2015}
\jname{Astronomy in Focus, Volume~1} 
\editors{Piero Benvenuti, ed.}
\begin{document}

\maketitle

\begin{abstract}
The study of galaxy mergers and supermassive binary black holes (SMBBHs) is central to our understanding
of the galaxy and black hole assembly and (co-)evolution at the epoch of structure formation and 
throughout cosmic history.  
Galaxy mergers are the sites of major accretion episodes, they power quasars, grow 
supermassive black holes (SMBHs), and drive SMBH-host scaling relations.   
The coalescing SMBBHs at their centers are the loudest sources of gravitational waves (GWs)
in the universe, and the subsequent GW recoil has a variety of potential astrophysical
implications which are still under exploration.  
Future GW astronomy will open a completely new window on structure formation and galaxy mergers, 
including the direct detection of coalescing SMBBHs, high-precision measurements of
their masses and spins, and constraints on BH formation and evolution in the high-redshift 
universe.   

\end{abstract}


\section{Introduction}\label{sec:intro}
 
Electromagnetic observations have provided us with numerous insights into the cosmic
growth history of black holes and the (co-)evolution with their host galaxies. 
Important advances of recent years 
include the discovery of quasars 
beyond redshift $z=6.5$, population studies in deep and wide field surveys constraining
the luminosity function and BH mass function out to high redshift, detailed studies
of nearby galaxy mergers, and the emergence of binary active galactic nuclei (AGN),
candidate SMBBHs, and candidate recoiling SMBHs.   
These have been accompanied by substantial progress on the theory side, including 
cosmological simulations of large-scale structure, galaxy merger simulations at super-high
resolution and approaching the ``final parsec'', breakthroughs in numerical relativity
enabling simulations of SMBBH coalescences, and intense ongoing investigations 
of potential electromagnetic signals quasi-simultaneous with binary coalescence.    

Despite this progress, a number of important questions related to the growth and cosmological
evolution of BHs remain partially unanswered: 

{\begin{itemize}

\item (1) How and when did the first BHs form ?  

{\item {(2) How and when did they grow and evolve ?{\footnote{What is the relative contribution of gas accretion,
BH--BH mergers, and stellar tidal capture/disruption ? What are the timescales, rates, efficiencies,
and trigger mechanisms ?}} }} 

\item (3) How do they evolve with respect to their host galaxies ?

\item (4) How often do binary SMBHs coalesce ?

\item (5) What is the BH spin evolution ?

\item (6) How frequent is GW recoil, and what are its astrophysical implications ?     

\end{itemize} } 

In the future, all of these questions can be independently addressed by GW
observations from space and using pulsar timing arrays (Sect. 6).

\section{Early SMBH growth and the highest-redshift quasars} 

Particularly tight constraints on the {\em early} BH growth have emerged from  
the discovery of luminous high-redshift quasars. 
At the time of writing, seven were known beyond 
redshift $z > 6.5$ (Mortlock et al. 2011, de Rosa et al. 2014, 
Venemans et al. 2015), including the one  at highest 
redshift of $z=7.1$ (Mortlock et al. 2011), 
corresponding to a cosmic age of only 700 Myr. 

Black hole masses of objects seen at those redshifts are high (Fig. 1), on the order
of $\sim 10^{9}$ M$_{\odot}$ and beyond,
reaching $\sim 10^{10}$ M$_{\odot}$ in two cases (at $z=6.3$, Wu et al. 2015;
and $z=5.4$, Wang et al. 2015), implying that very massive BHs were in place early in the universe. 
They do not grow much beyond that anywhere in
the cosmos (Fig. 1) as mass estimates of the most massive BHs amount to 
a few\,10$^{10}$ M$_{\odot}$ (e.g., Shen et al. 2011, but see also Brockamp et al. 2015). 

How fast can massive BHs grow early in the universe? 
BHs accreting at the Eddington limit, 

\begin{equation}
L_{\rm Edd} = {4 \pi G M m_{\rm p} c \over{\sigma_{\rm T}}} 
    \simeq 10^{38}\,{\bigg{(}}{M\over {M_\odot}}{\bigg{)}} {\rm erg/s}\,,
\end{equation}

\noindent with accretion luminosity $L$=$\eta \dot{M} c^2$, 
where $\eta$ is the radiative efficiency,
grow exponentially, as 

\begin{equation}
M = M_{\rm init}\,e^{t/\tau_{\rm salpeter}}\,~{\rm with}\,~\tau_{\rm salpeter} = 4.5 \,10^7 (\eta/0.1)~ {\rm yr}\,.   
\end{equation}
  
The growth timescale depends on the radiative efficiency $\eta$ and therefore on BH spin. 
Higher BH spin implies last stable orbits closer in, higher radiative efficiency, and therefore
less rapid BH growth. For instance, it takes $\sim$ 2 Gyr for a BH to grow up to $10^9$ M$_{\odot}$
if it is rapidly spinning ($\eta$=0.3). 
Even if BHs start accreting with low spin, the accretion process itself 
(if from a coherent, long-lived disk), will rapidly spin
up the hole (Volonteri et al. 2013).  
The detection of high-redshift quasars with SMBH masses as high as $\sim10^{10} M_{\odot}$ at $z=6.3$ 
(Wu et al. 2015) then implies that there is not enough time for them to grow via Eddington-limited accretion from
low-mass seed BHs. Supercritical accretion and/or massive seeds are possible solutions (e.g., Volonteri et al. 2015).    
 
With future GW observations (Sect. 6) we will be sensitive not only to the most massive black holes,
but also to high-redshift mergers of black holes as much as a million times smaller.  
These observations will motivate new questions 
not only about how the biggest black holes grew but about the yet unseen broader 
population of objects, including the precursors of the typical SMBHs observed 
in the present day universe.

\begin{figure}[t]
\begin{center}
 \includegraphics[width=11cm]{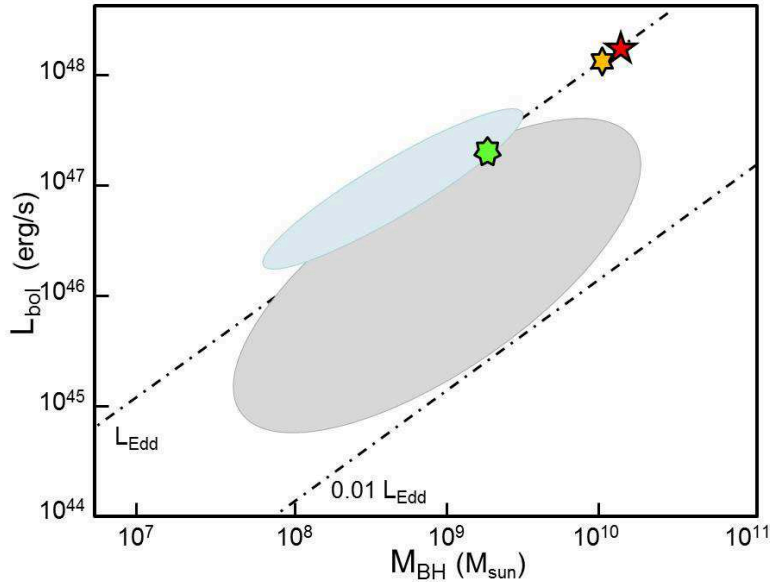}
 \caption{BH masses of high-redshift quasars (star symbols, from left to right:
  ULASJ1120+0641 at $z=7.1$, J0306+1853 at $z=5.4$, and SDSSJ0100+2802 at $z=6.3$) in comparison
  with other $z>6$ quasars (small grey ellipse), and the area populated by a large number of
  lower-redshift SDSS quasars
   (large grey ellipse; adopted from Wang et al. 2015).
}
   \label{fig1}
\end{center}
\end{figure}

\section{Galaxy mergers}

Gas-rich galaxy mergers trigger quasar activity, provide the fuel source for BH growth,
are the sites of feedback processes, and likely drive the BH-host scaling relations.  

\subsection{Triggering of quasar activity}

The idea that luminous quasars are activated by galaxy interactions
came up early, in the seventies (e.g., Stockton \& MacKenty 1983). Since then, numerous
studies have been carried out to address whether all AGN activity is triggered by mergers.
Based on a recent compilation of results from several IR, optical and X-ray surveys up to $z\sim3$,
Treister et al. (2012) concluded, that the fraction of AGN in mergers is a strong function
of, and increases with, AGN luminosity. The majority of the most luminous quasars all reside 
in major mergers{\footnote{however, not all samples show this trend (e.g., Villforth et al. 2014)}}.  
Lower-luminosity AGN, on the other hand, are likely triggered by secular processes like
bar-driven inflows, stochastic cloud accretion events, minor mergers and perhaps stellar
captures (Hopkins \& Hernquist 2009). 
These differences may also explain cosmic downsizing (review by Brandt \& Hasinger 2005). 
While the low-mass, low-luminosity AGN make the majority in {\em number}, a
large part of the total BH {\em mass growth} likely occurrs in the most luminous 
quasars (e.g., Treister et al. 2012). 
Runaway BH growth in these gas-rich environments is plausibly prevented by feedback processes,
which have also been invoked to explain the scaling relations 
between BHs and their host galaxies.  
  
\subsection{BH-host scaling relations} 

The mass of the central SMBH correlates tightly with a number of host galaxy
properties including stellar velocity dispersion and total bulge mass{\footnote{a few 
extreme outliers of very massive BHs in low-mass host
galaxies have been reported recently (e.g., van den Bosch et al. 2012, Trakhtenbrot et al. 2015)}} 
(review by Graham 2015). 
Galaxy mergers are thought to be the  main driver behind these relations, either by merging
repeatedly with each other (Jahnke \& Maccio 2011), or by triggering feedback processes such that the
radiation from the accretion disk heats the ambient gas and drives strong outflows until -- at
a critical BH mass -- the remaining gas is expelled which then terminates further BH growth 
and fixes the host properties (e.g., Hopkins et al. 2006).  

\section{Evolution of galaxy mergers and SMBBHs} 

Galaxies merge frequently throughout cosmic times. Whenever both galaxies harbor SMBHs at their centers,
the formation of a binary SMBH is inevitable. The merger evolves in several stages (Begelman et al. 1980;
our Fig. 2). (1) The early stages of galaxy merging are driven by dynamical friction. (2) At close separations,
on the order of parsecs, the two SMBHs form a bound pair. The further shrinkage of their orbit then depends
on the efficiency of interactions with gas and stars in carrying away energy and angular momentum; known
as the ``final parsec problem'' (review
by Colpi 2014). (3) At separations well below a parsec, emission of GWs becomes efficient leading
to further orbital shrinkage and final coalescence. This GW-driven regime can be thought of as 
proceeding in three stages; the inspiral phase, dynamical merger, and final ringdown, which emit 
characteristic GW radiation (e.g., Hughes 2002; review by Centrella et al. 2010).  
(4) Depending on the orbital configuration including
masses and spins of the binary, after coalescence the newly formed 
single SMBH then recoils with a velocity as high 
as $\sim$ 5000 km/s (Lousto \& Zlochower 2011; review by Sperhake 2015) 
but typically much lower.   

\begin{figure}[t]
\begin{center}
 \includegraphics[width=12cm]{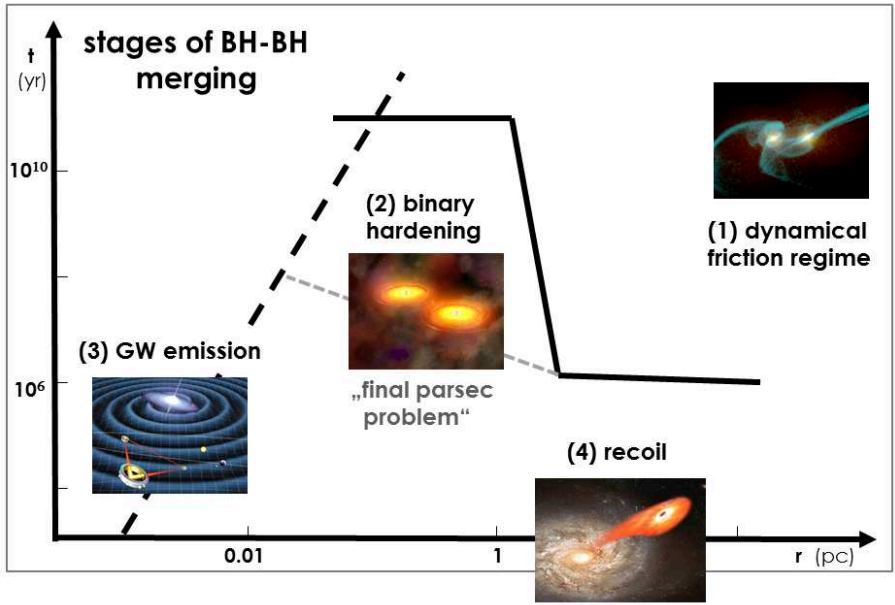}
 \caption{Stages of galaxy and SMBBH merging. 
}
   \label{fig1}
\end{center}
\end{figure}

\section{Observations of (U)LIRGs, binary AGN and SMBBHs}

Prime examples of gas-rich galaxy mergers are luminous and ultra-luminous infrared galaxies
[(U)LIRGs; Sanders et al. 1988]. These allow us to study the stages of galaxy merging 
in the nearby universe, and are more abundant in the 
higher-redshift universe, where they become
a major galaxy population. Based on an analysis of {\itshape{HST}} images, 
Kartaltepe et al. (2012) concluded, that at least 70\% of ULIRGs
at redshift $z\sim2$ reside in interacting systems.    
Nearby (U)LIRGs provide us with excellent laboratories for 
studying the physics and evolution of galaxy
mergers, and the triggering of AGN activity at their centers (e.g., Iwasawa et al. 2011).
One of the first observed with high-resolution X-ray imaging spectroscopy with {\it{Chandra}}
was the nearby ULIRG NGC\,6240, which let to the detection of a pair of X-ray luminous
AGN at its center (Komossa et al. 2003).   

Meanwhile, more cases of AGN pairs have emerged, based on X-ray, radio and optical imaging
spectroscopy. The majority of these is found in galaxy pairs in relatively early stages
of their evolution  (see Fig. 2 of Deane et al. 2014 for a compilation of wide 
systems at $\leq$ 10 kpc spatial separation, as well as recent work by
Comerford et al. 2015, Fu et al. 2015, De Rosa et al. 2015, and references
therein). Only 3--5 systems are currently known
which host binary AGN below $\sim$1 kpc spatial separation, which are located in single galaxies
or advanced mergers. Besides NGC\,6240, these are J0402+379 in the radio (at only 7pc projected
separation; Rodriguez et al. 2006, Burke-Spolaor 2011), 
SDSSJ1323--0159 in the optical (Woo et al. 2014), NGC\,3393 in
X-rays (Fabbiano et al. 2011; see Koss et al. 2015 for a different interpretation), 
and SDSSJ1502+1115 (Deane et al. 2014;
see Wrobel et al. 2014 for a different interpretation).  
All of these nearby pairs, upon their final coalescence, would be easily detectable in GWs with
a {\it{LISA}}-type mission (e.g., Fig. 1 of Colpi 2014).  
At present, radio-VLBI observations provide us with the most powerful technique of resolving SMBH pairs
at small angular separation. 

The most compact systems, supermassive binary BHs, can no longer 
be spatially resolved with current techniques, and we therefore
rely on indirect methods when searching for them electromagnetically. Most search strategies 
are based on signs of semi-periodicity, for instance in lightcurves{\footnote{the blazar OJ\,287
with its $\sim$ 12 yr optical periodicity is one of the best-studied cases (e.g., Valtonen et al. 2012)}} 
or in the spatial structures of radio jets. Others include double-peaked emission lines
or unusual spectral energy distributions (SEDs) at the time when the advanced binary
has opened a gap in the inner accretion disk (see Komossa \& Zensus 2016
for a review of signatures and pre-2015 SMBBH candidates).  
Several recent candidates have emerged from large-sky surveys and long-term monitoring programs. 
PG\,1302--102 (Graham et al. 2015, D'Orazio et al. 2015, see also Kun et al. 2015), 
PSOJ334.2028+01.4075 (Liu et al. 2015), and 
PG\,1553+113 (Ackermann et al. 2015)
all show pronounced semi-periodic lightcurve variability; while Mrk\,231 (Yan et al. 2015)
exhibits a UV-dim SED characteristic
of a binary-driven gap in the accretion disk.  
 
All of these detection methods of SMBBHs require at least one BH to be active. In order to
trace the population of inactive binaries, Liu et al. (2009) proposed to take advantage of epochs of
temporary activity in form of accretion flares from tidally disrupted stars in these systems. 
As the second BH temporarily interrupts the stellar debris stream on the primary, it causes
characteristic dips and recoveries in the tidal disruption lightcurve. This signature has
been identified in the lightcurve of SDSSJ120136.02+300305.5, which is consistent with 
the presence of a sub-milliparsec SMBBH (Liu et al. 2014).          

Finally, post-coalescence candidates include radio galaxies with characteristic structures [double-doubles
or X-shaped systems (e.g., Liu et al. 2003, Roberts et al. 2015), interpreted as evidence for the interruption and
re-start of accretion activity during coalescence, and/or BH spin flips],   
and recoiling black holes of which a few candidates have been identified 
in recent years (review by Komossa 2012).

\section{What do GWs tell us about structure formation ?}

All of the questions raised in Sect. 1 can efficiently be addressed with GW astronomy with
space-based observatories and pulsar timing arrays (e.g., Hughes 2002, Menou 2003, Centrella 2003,
Sesana 2013, McWilliams et al. 2014, Colpi 2014, Barausse et al. 2015), 
in a completely independent
and sometimes unique way. 

We do not yet have any direct observations of the first seed BHs. They may form as light
seeds from the collapse of the first massive stars, or as massive seeds through direct collapse
of large gas clouds. In the near future, we will not be able to observe 
this regime directly electromagnetically, except perhaps through high-redshift GRBs.  
In GWs, and with an {\it{eLISA}}-type mission, low-mass BHs would be
observable out to redshifts $z < 15$ (Colpi 2014); in a crucial regime of galaxy
assembly and seed BH formation and growth.  

Further, GWs allow us to measure luminosity distances, and then serve as 
standard candles (standard sirens; Schutz 1986, review by Barausse et al. 2015){\footnote{if potential problems
posed by weak lensing can be overcome}}.

While current searches for tightly bound SMBBHs continue to be challenging, 
and we still lack ``smoking gun'' signatures for incontrovertible 
identifications of such binaries, GW observations can provide a strongly complementary view.  
Where gravity is strong enough to produce observable GWs, 
it is expected to dominate the process, leading to clear GW signatures with very 
straightforward physical interpretations.  
Further, if multimessenger observations are possible then a few GW observations 
may be leveraged to eventually hone our interpretation of more numerous electromagnetic 
binary candidates. 

In particular, GWs from coalescing binaries enable high-precision measurements  of BH masses and spins.
Measurements of coalescence rates (including extreme 
mass-ratio inspirals) inform us
about the merger history of galaxies and constrain the accretion history (from spin measurements). 
Pulsar timing arrays have just started to place constraints on 
galaxy merger history from limits on the
stochastic GW background (e.g., Zhu et al. 2014, Arzoumanian et al. 2015,
Lentati et al. 2015, 
Shannon et al. 2015).  
GW astronomy therefore provides us with unique tracers of black hole assembly and growth,
including new constraints on the importance of accretion, merging and stellar captures in growing
black holes, and on the BH spin history.

%


\end{document}